\documentclass[a4paper,11pt]{article}
\pdfoutput=1 

\usepackage{jinstpub} 

\title{Imaging by muons and their induced secondary particles -- a novel technique}

\author[a,1]{G. Galg\'oczi,\note{Corresponding author.}}
\author[b]{D. Mrdja,}
\author[b]{I. Bikit,}
\author[b]{K. Bikit,}
\author[b]{J. Slivka,}
\author[b]{J. Hansman,}
\author[a]{L. Ol\'ah,}
\author[a]{G. Hamar}
\author[a]{and D. Varga}

\affiliation[a]{Wigner Research Centre for Physics,\\1121 Budapest, Konkoly-Thege Mikl\'os\'ut 29-33., Budapest, Hungary}
\affiliation[b]{University of Novi Sad, Faculty of Sciences, Department of Physics,\\ Trg Dositeja Obradovica 4, 21000, Novi Sad, Serbia }

\emailAdd{galgoczi@caesar.elte.hu}

\abstract{Muography is a well estabilished method to obtain 3D images of large objects (e.g. volcanoes and large buildings) without any additional particle source, taking advantage of the presence of cosmic muons. The underlying principle of muography is the measurement of individual muon tracks and the determination of their absorption or scattering. These processes depend on the material that they have travelled through. The novel method discussed is based on the measurement of the muon tracks and of the corresponding particles those were produced by the muons themselves in the investigated target.

As muons pass through matter they interact with matter by ionization, bremsstrahlung, pair production and nuclear interactions. Our experimental setup is designed in a way to measure both the primary muons and the created secondaries (mostly electrons and gammas). The tracks of the muons are determined by a special kind of Multi-Wire Proportional Chambers (MWPC) called CCC (Close Cathode Chamber). The secondary particles produced in the target are measured by four plastic scintillators placed around the target. The CCC chambers and the scintillators are used in coincidence in order to gather data about muons those passed through the target. As cross sections of the described processes vary by the density and the atomic number of materials this technique could be used to investigate the material content of the target.}

\keywords{simulations, concepts, interaction of radiation with matter}

\proceeding{15$^{\text{th}}$ Topical Seminar on Innovative Particle and Radiation Detectors\\
  14-17 October 2019\\
  Siena, Italy}

\begin{document}
\maketitle
\flushbottom

\section{Introduction}
\label{sec:intro}

For most particle physics experiments cosmic muons are a background which needs to be taken into account. Although in the last 50 years the presensce of cosmic muons have been used also as a tool to obtain 3D images of large objects, e.g. pyramides and vulcanoes \cite{1,2,3}. In our experiment we have the aim to prove that secondary particles created by cosmic muons upon crossing matter could be used to distinguish different materials \cite{4,45}. This could be useful for small scale objects which other non invasive methods (e.g. X-ray) are not able to penetrate, for example meteorites. Our experimental setup consists of a set of trackers to identify cosmic muons and large area scintillators to detect the secondary gammas and electrons created by the interaction of the muons with matter.

\section{Secondaries produced by cosmic muons}
\label{sec:secs}

The main difference between our experimental setup and the currently existing projects observing muons with cosmic origin is the fact that we take advantage of detecting the secondary particles produced by the muon interacting with the target material. By measuring the number, the energy and the type of the secondary particles we can gain information about the material which the muon passed through. Muons interact with matter by ionization, pair production, Bremsstrahlung and nuclear interaction.

\begin{figure}[h!]
\centering 
\qquad
\includegraphics[width=.45\textwidth,origin=c]{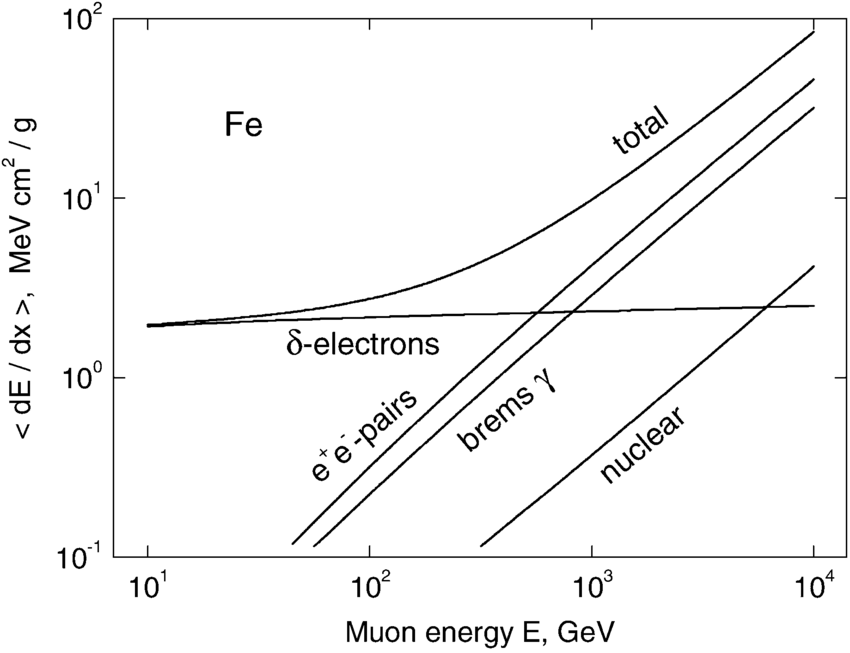}
\caption{\label{fig:interactions} Energy loss of muons due to different interactions inside iron \cite{a}.}
\end{figure}

\newpage

In fig. $\ref{fig:interactions}$ the importance of each interaction can be understood as a function of muon energy. In our case for cosmic muons ionization, pair production, Bremsstrahlung are also important. For low-Z materials delta-ray production is more relevant. On the other hand, for high-Z materials Bremsstrahlung yields to the majority of the detections.

\section{MUon CAmera (MUCA)}
\label{sec:muca}

The MUon CAmera (MUCA) \cite{b,c} consists of a tracker system and a set of scintillators. The previous is on the top and bottom of the target volume and the latter is around it. The tracker system consists of a set of Close Cathode Chambers (CCCs) \cite{cd} and its aim is to detect cosmic muons passing through the target volume. The scintillators placed around the target cover 75 \% of the solid angle in order to detect the secondary particles produced by muons inside the target volume. The scintillators and the tracker system are read out in coincidence.

\begin{figure}[h!]
\centering 
\qquad
\includegraphics[width=.48\textwidth,origin=c]{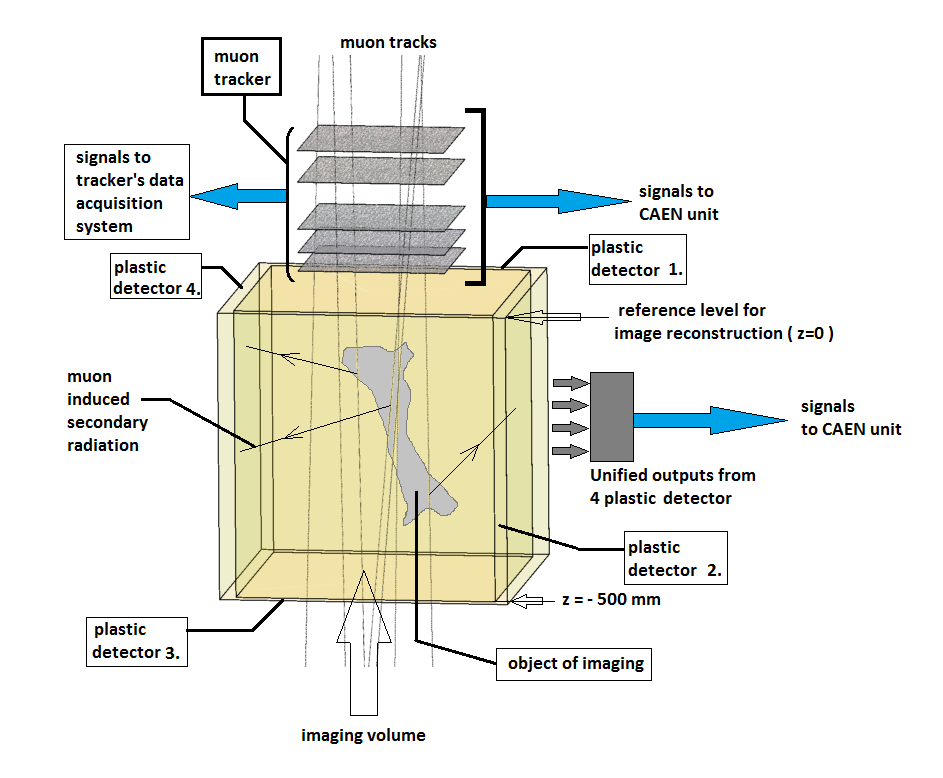}
\caption{\label{fig:bonemuca} The schematic drawing of the MUCA system examining a bone. \cite{b} Trackers above the target identify muons and the scintillators on the sides detect the secondaries produced inside the target.}
\end{figure}

Using coinicidence with the 200 ns time resolution of the detectors lowers the gamma background to almost zero. The required acquisition time depends on the density and atomic number of the target material. It varies from hours to days.

\begin{figure}[h!]
\centering 
\qquad
\includegraphics[width=.95\textwidth,origin=c]{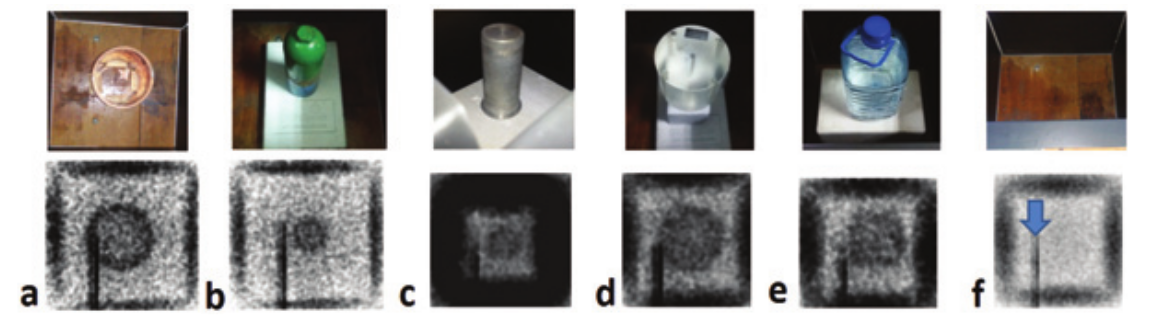}
\caption{\label{fig:difftargets} (a) Cu cylinder (b) Iron cap (c) Aluminum cap (d) Headphantom (e) Water in plastic bottle (f) Background image (empty imaging volume) an arrow appeared due to electronic noise in the tracker's layer. Acquisition time $\sim$ 1 day \cite{b}.}
\end{figure}

\section{Geant4 simulations}

A wide variety of materials were investigated by the MUCA system \cite{b} including materials with high and low atomic number and also with very different densities. A few examples can be seen in fig. $\ref{fig:difftargets}$. The energy spectrum of the secondary particles produced by the cosmic muons  currently measured by the four scintillators surrounding the target. A series of Geant4 \cite{5} simulations were developed to understand what particles are produced in different materials and also to take into account the energy deposition and the optical light propagation inside the scintillators.

\begin{figure}[htbp]
\centering 
\qquad
\includegraphics[width=.6\textwidth,origin=c]{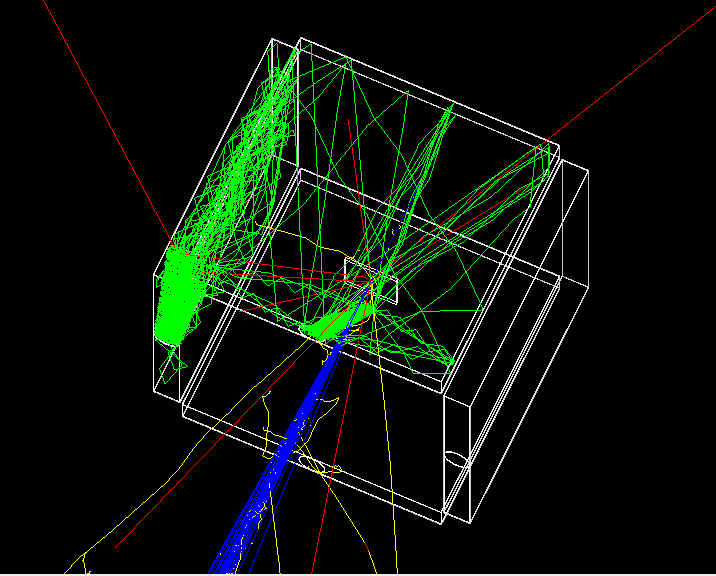}
\caption{\label{fig:g4} Geant4 simulation of a 10 x 10 x 0.5 cm$^{3}$ lead brick in the middle hit by 50 muons with energy of 1 GeV.}
\end{figure}

 In fig. $\ref{fig:g4}$ the secondary particles produced inside the target lead brick and the optical photons produced inside the scintillators can be seen. Yellow tracks represent electrons, the red ones gammas. The green ones stand for optical photons and the blue ones are the primary muons. Only those optical photons are plotted, which were detected by the PMTs.

\section{Results}
\label{sec:res}

\subsection{Secondaries created and exiting the target volume}

\begin{figure}[h!]
\centering 
\includegraphics[width=.45\textwidth,clip]{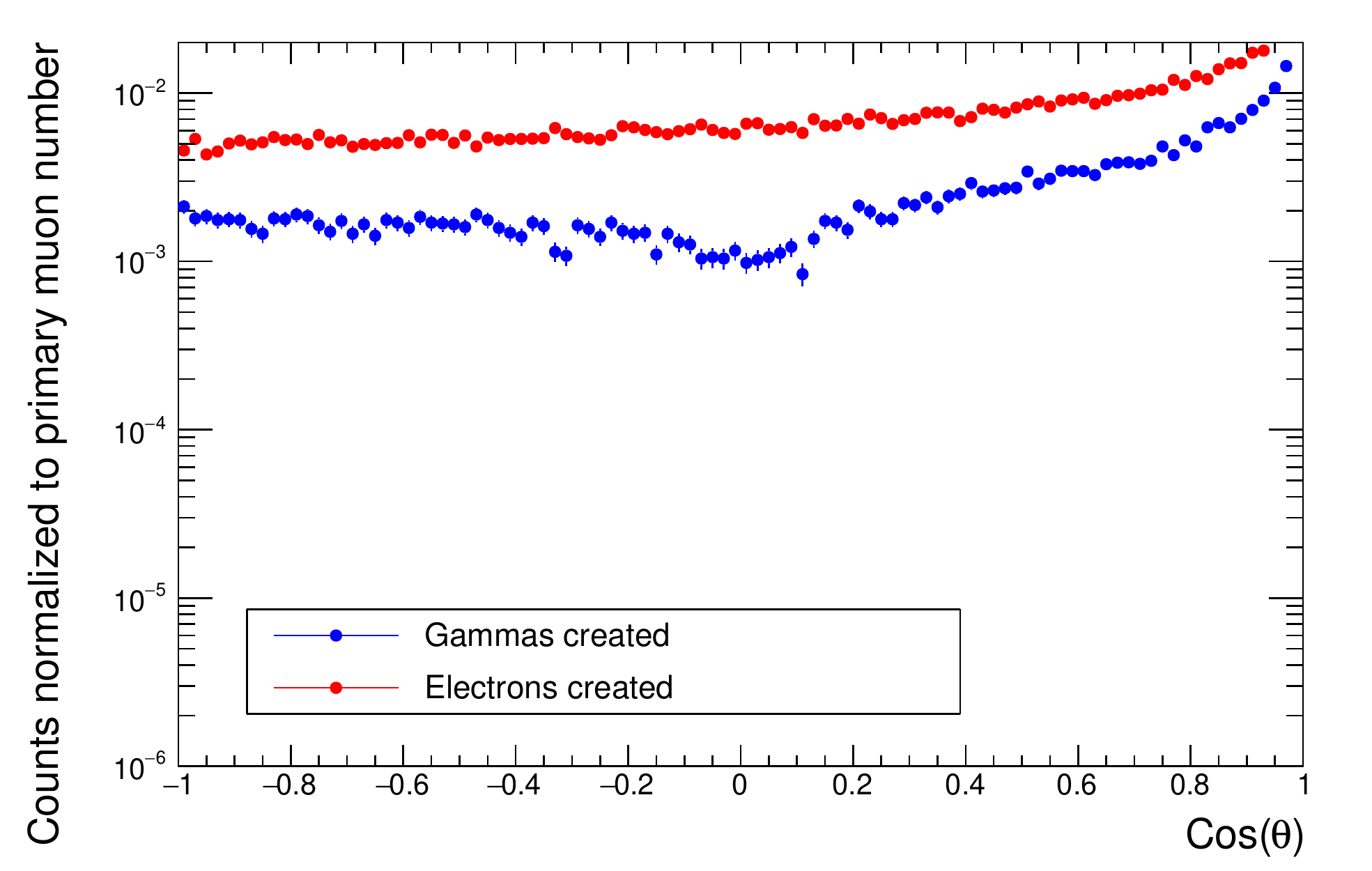}
\qquad
\includegraphics[width=.45\textwidth,origin=c]{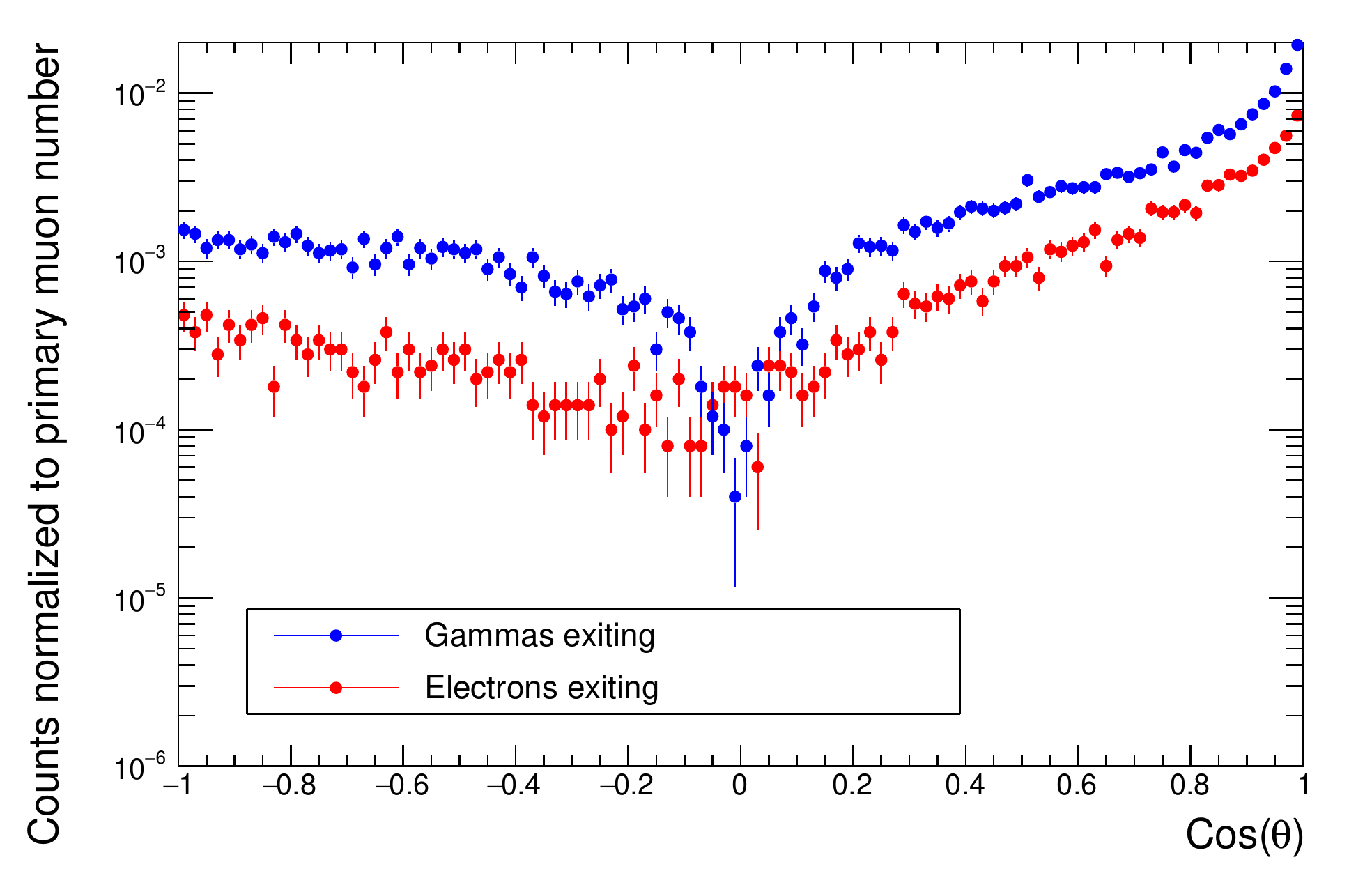}
\caption{\label{fig:azimuthal} The polar angle distribution of all created secondary particles (left) and the polar angle distribution of secondaries exiting a lead target with a size of 10 x 10 x 0.5 cm$^{3}$  (right).}
\end{figure}

From an experimental point of view, the directional momentum distribution of the created secondary particles is extremely relevant as the flux of cosmic muons is low. Therefore to keep the required time of measurement with the MUCA system within realistic values one needs to capture as many secondares as possible. In fig. $\ref{fig:azimuthal}$ the polar angle of the created electrons and gammas is plotted. Secondaries favour the forward direction due to the relativistic energy of the cosmic muons. Secondaries exiting the target favour the forward direction even more as the high energy particles which have enough energy to exit the material especially favour the forward direction. (Also there is less material to be crossed in this direction.)

\begin{figure}[h!]
\centering 
\includegraphics[width=.45\textwidth,clip]{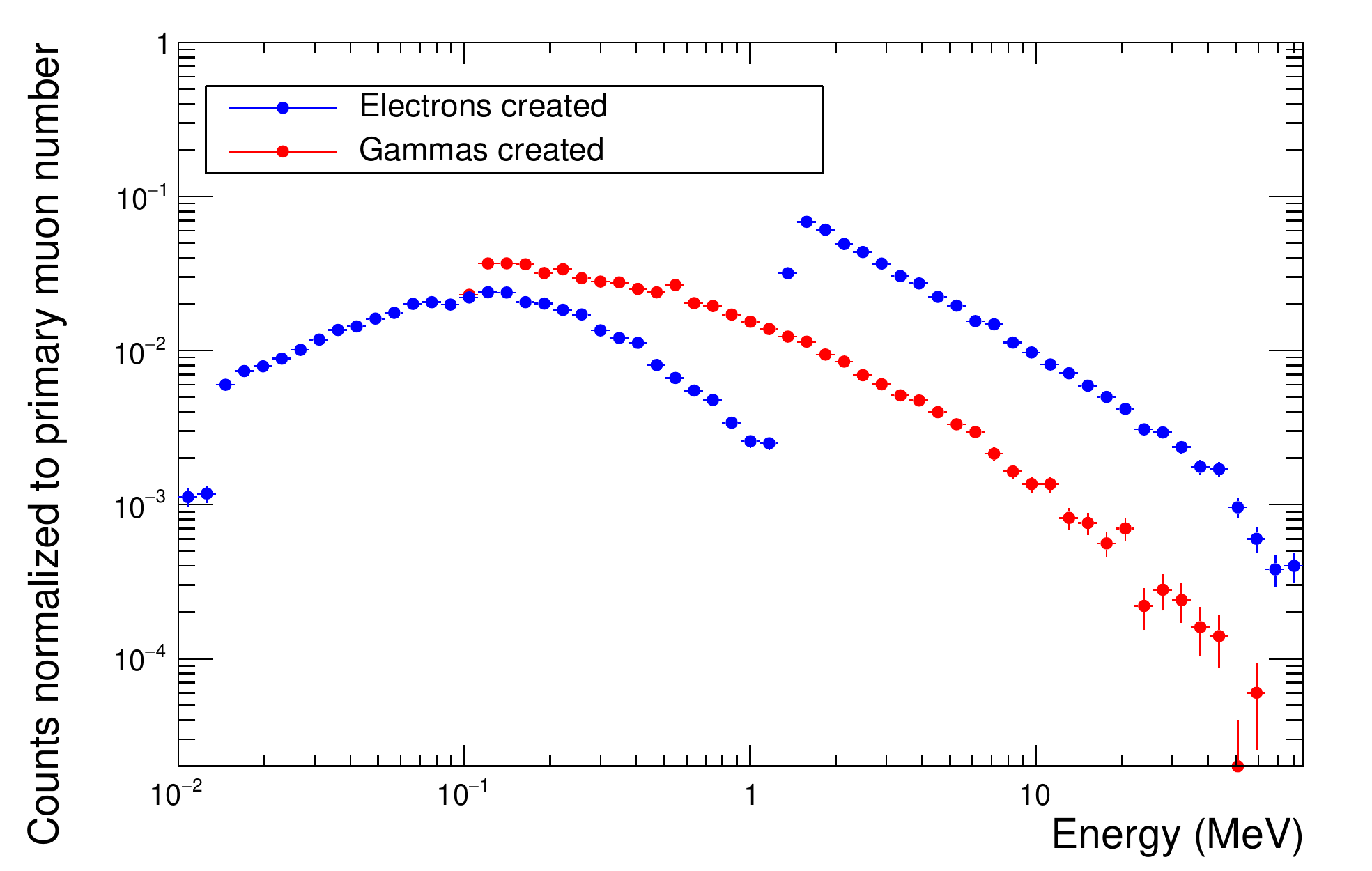}
\qquad
\includegraphics[width=.45\textwidth,origin=c]{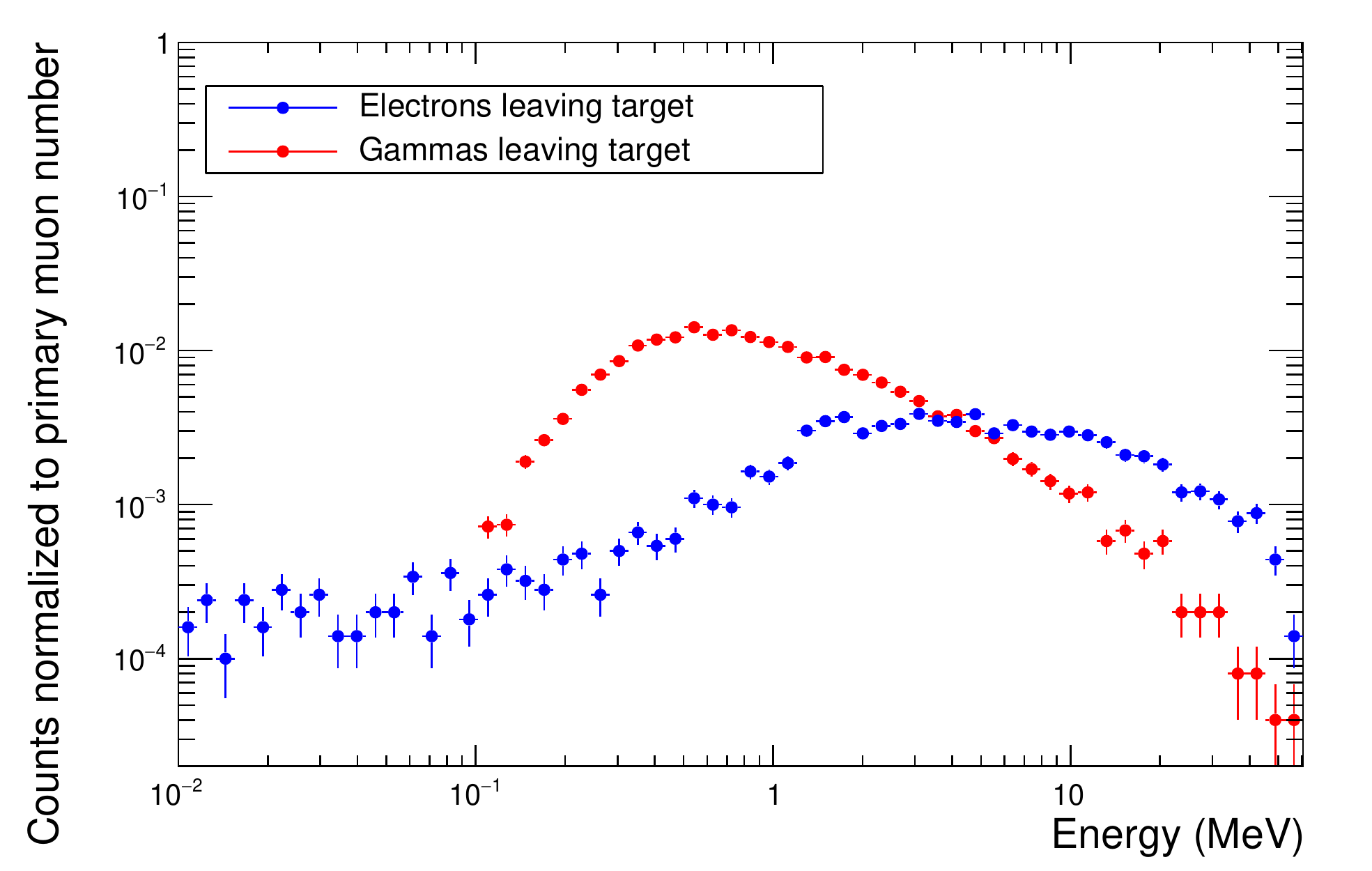}
\caption{\label{fig:energy} The energy distribution of all created secondary particles (left) and the energy distribution of secondaries exiting a lead target with a size of 10 x 10 x 0.5 cm$^{3}$(right).}
\end{figure}

The spectrum of the produced secondary particles can be seen in fig. $\ref{fig:energy}$ for muon energy of 1 GeV and a lead target. Note that the drop in the electron and the gamma spectrum is due to the production cut in the simulation in order to keep computational time realistic. This specific cut was chosen to be lower than the required energy for electrons to exit the target and to arrive to the scintillators. This is the energy which is needed for the detection of electrons. In fig. $\ref{fig:energy}$ shows the spectra of particles leaving the same target. Secondary electrons lose much more energy therefore they are more surpassed. 

\subsection{Distinguishing materials by the produced secondaries}

Our simulations show that a simple way to distinguish materials is to measure the ratio and number of gamma particles and electrons detected. In table $\ref{tab:1}$ the likeliness of secondary particles exiting with the corresponding energy treshold is shown. The target was chosen to be 10x10x0.5 cm$^3$. This energy treshold was chosen to be 0.1 MeV for gammas and 1 MeV for electrons as this is roughly the minimal energy that is required to arrive to the detectors and be detected. The ratio of gammas and electrons varies more than 2 orders of magnitude for different materials investigated. 

\begin{table}[h!]
\begin{center}

\begin{tabular}{|l|llll|}
\hline
 & Lead & Copper & Water & Polystyrene \\\hline
$\gamma$, E $\geq$ 0.1 MeV & 20.7 \% & 10.6 \% & 0.14 \% &  0.14 \%  \\
e$^{-}$, E $\geq$ 1 MeV & 6.68 \% & 13.1 \% &  6.08 \% & 4.7 \% \\\hline
  
\end{tabular}

\caption{\label{tab:1}Likeliness of secondary particles leaving the target with given energy treshold which corresponds to the required energy to be detected.}
\end{center}
\end{table}

In the case of lead every fifth muon produces a gamma with more energy than 100 keV and for polystyrene only one from one thousand muon produce a gamma with at least 100 keV. It is interesting to mention that the number of gammas produced in low Z material are similar but the number of electrons is different. For water there is 25\% more electrons produced than for polystyrene.

\end{document}